\documentclass{article}
\usepackage{spconf,amsmath,amssymb,graphicx, url}

\title{Generating Novel and Realistic Speakers for Voice Conversion}
\name{Meiying Melissa Chen, Zhenyu Wang and Zhiyao Duan}
\address{Department of Electrical and Computer Engineering, University of Rochester, Rochester NY, USA\\
meiying.chen@rochester.edu}
\begin{document}
%
\maketitle
\begin{abstract}
Voice conversion models modify timbre while preserving paralinguistic features, enabling applications like dubbing and identity protection. However, most VC systems require access to target utterances, limiting their use when target data is unavailable or when users desire conversion to entirely novel, unseen voices. To address this, we introduce a lightweight method SpeakerVAE to generate novel speakers for VC. Our approach uses a deep hierarchical variational autoencoder to model the speaker timbre space. By sampling from the trained model, we generate novel speaker representations for voice synthesis in a VC pipeline. The proposed method is a flexible plug-in module compatible with various VC models, without co-training or fine-tuning of the base VC system. We evaluated our approach with state-of-the-art VC models: FACodec and CosyVoice2. The results demonstrate that our method successfully generates novel, unseen speakers with quality comparable to that of the training speakers.

\end{abstract}
\section{Introduction}
\label{sec:int}
Voice conversion (VC) models offer the capability to transform the timbre of a source speaker's voice to that of a target speaker \cite{ju2024naturalspeech, du2024cosyvoice, chen2022controlvc}. This is highly valuable in various applications. For instance, a single voice actor could dub an entire film using multiple distinct voices and performance styles. Another significant use case involves individuals wishing to alter their voice to protect their identity.
Consequently, users often require not only novel, high-quality voices that are distinctive and free from legal restrictions, but also a diverse range of these generated voices to select from.

However, current voice conversion models typically depend on the access to pre-recorded target speaker utterances \cite{cao2024neuralvc, baade2024neural, zhang2025vevo}. Generally, two main approaches exist: disentanglement-based methods and large language model (LLM) based systems. Disentanglement-based approaches \cite{chen2022controlvc, yao2024promptvc, yang2024streamvc, wang2025spark}, represent speech through separate embeddings for its distinct components, such as content, speaker identity, and pitch. During conversion, these methods require extracting speaker embeddings from target recordings to serve as input, while other components like content and prosody are retained from the source speaker. LLM-based systems \cite{du2024cosyvoice, chen2024f5}, on the other hand, utilize acoustic features like mel-spectrograms or speech tokens from the target voice as prompts. These prompts are often concatenated with other conditioning information, such as content representations, and then directly input into the speech-LLM to generate acoustic tokens for a subsequent vocoder. 
Both methods, however, assume that target utterances exist and thus lack the ability to generate new speakers.

Some recent models use interpolation to generate new speaker embeddings by combining existing ones \cite{chattts2024}. However, these methods often do not ensure that the generated speakers are sufficiently novel or of high audio quality. While certain text-to-speech (TTS) models can generate novel speakers using label inputs, like sex and age, this technique has not yet been applied to voice conversion \cite{tacospawn, shi2023voicelens}. Moreover, in many such studies, speaker generation is a side feature rather than the primary research focus. As a result, existing methods lack systematic methodology to evaluate the fidelity and novelty of these generated speakers, specifically, whether the proposed generation methods truly produce speakers distinct from the training set and whether these new speakers generate high audio quality. 

In this paper, we address these limitations by proposing a novel speaker generation method named SpeakerVAE to create new, natural-sounding speakers specifically for voice conversion applications. Our approach utilizes a deep hierarchical variational autoencoder \cite{vahdat2020nvae} to model the speaker space learned by the speaker module of a pre-trained voice conversion system. By sampling from this learned VAE space, we can generate diverse and unique speaker representations.

The proposed system offers two key advantages:

\textbf{Efficiency}: It requires minimal training, focusing only on the VAE module, with no retraining needed for the existing voice conversion model.

\textbf{Flexibility}: The method is adaptable and can be applied to various voice conversion systems and different distributions of speaker embedding spaces, whether they are regulated speaker verification embeddings or unregulated style vectors learned jointly with a synthesis system.

Additionally, we propose a systematic approach to evaluate the naturalness and range of the generated speakers. This evaluation framework aims to ensure that the newly generated speakers not only exhibit high generation quality but also possess statistical features comparable to those observed in the training dataset. The subsequent sections will detail the architecture of our proposed SpeakerVAE, the generation process, and the comprehensive evaluation results.

\section{Speaker Generation with NVAE}
\label{sec:met}
The key idea of our pseudo speaker generation approach is to learn a generative model from speaker embeddings extracted from utterances of a training dataset with many different real speakers. After learning this generative model, one can then sample a pseudo speaker embedding and use it as the target speaker embedding in voice conversion applications. Ideally, the sampled pseudo speaker embedding does not collide with any real speaker embeddings in the training set, but its distribution follows that of the training speaker embeddings, ensuring the naturalness and range of pseudo speakers.

There are many generative models to consider for this purpose. In this paper, we propose to use the Nouveau Variational Auto-Encoder (NVAE) \cite{vahdat2020nvae}, a deep hierarchical VAE.
We choose NVAE due to its ability to model complex, high-dimensional distributions through a hierarchical structure. Speaker embeddings, which encapsulate a diverse range of human vocal characteristics, inherently form such a complex space. These characteristics include variations in pitch, timbre, and other nuanced acoustic features that contribute to unique speaker identities. NVAE's hierarchical approach is particularly good at learning representations at different levels of abstraction from such data. Furthermore, NVAE is designed to avoid over-regularization and preserve fine-grained variability, which is critical for generating diverse yet natural-sounding voices.


The speaker embeddings modeled by NVAE do not need to follow a specific distribution. They can be extracted from speech utterances with speaker verification models such as ECAPA-TDNN \cite{desplanques2020ecapa}. They can also be extracted by speech encoders of voice conversion systems \cite{li2024gtr, li2023styletts}. We argue that the latter may contain richer information for rendering high-quality and natural speech utterances in voice conversion applications. 
In this work, we train NVAE with both first type speaker verification embeddings (used in CosyVoice2\cite{du2024cosyvoice}) and second type embeddings (used in FACodec \cite{ju2024naturalspeech}) to verify its performance.



\begin{figure*}[htbp]
    \centering
    \includegraphics[width=1.0\linewidth]{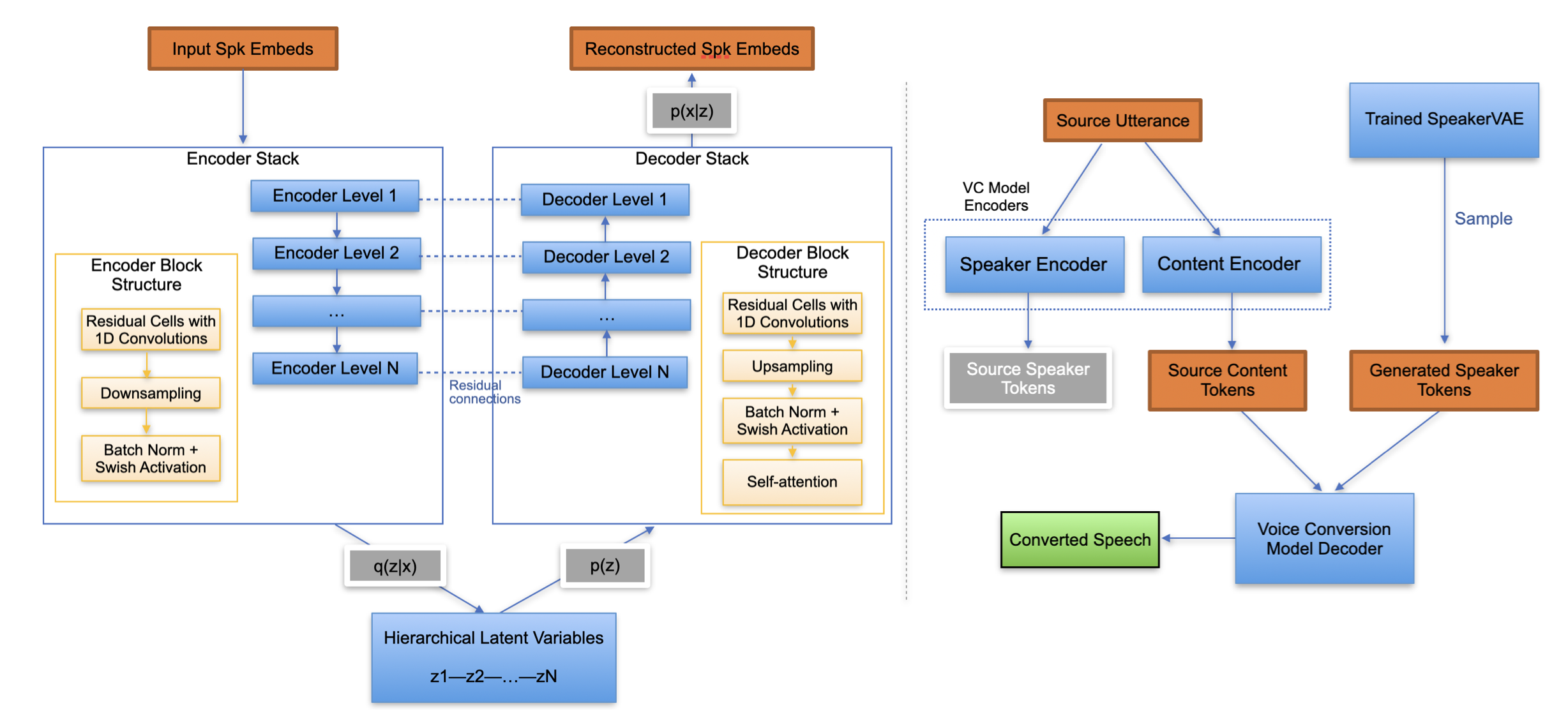}
    \caption{SpeakerVAE overview. Left side shows the architecture and training process. Right side depicts the inference pipeline.}
    \label{fig:system}
\end{figure*}

\subsection{NVAE Model Architecture}
The original NVAE architecture \cite{vahdat2020nvae} is a deep hierarchical VAE designed for high-quality image generation, leveraging a multi-scale latent space and residual cells to model complex pixel correlations. It features a sequence of latent variables $z = (z_1, \ldots, z_L)$, where $L$ is the total number of hierarchical latent groups and each $z_l$ represents the latent variables at level $l$. The generative model $p_\theta(x, z)$ (with $\theta$ denoting the decoder parameters and $x$ being the input data) follows a top-down process:
\begin{equation}
p_\theta(x, z) = p_\theta(x|z_L) \prod_{l=1}^{L} p_\theta(z_l|z_{<l}),
\end{equation}
where $p_\theta(z_1|z_{<l}) = p_\theta(z_1)$ forms the base case prior, $z_{<l} \equiv (z_1, \ldots, z_{l-1})$ represents all coarser-level latents, $p_\theta(x|z_L)$ is the observation model, and $p_\theta(z_l|z_{<l})$ is the conditional prior for $z_l$ given latents from coarser levels $z_{<l}$.

The encoder $q_\phi(z|x)$ (where $\phi$ represents the encoder parameters) uses a bottom-up inference process, where each $q_\phi(z_l|x, z_{>l})$ (the approximate posterior) conditions on $z_{>l} \equiv (z_{l+1}, \ldots, z_L)$  (the finer-level latents). NVAE stabilizes training via spectral regularization  and residual cells with skip connections. The model optimizes the Evidence Lower Bound (ELBO):
\begin{equation}\label{eq:elbo}
	\begin{split}
&\mathcal{L}(\theta, \phi; x) = \mathbb{E}_{q_\phi(z|x)} \left[ \log p_\theta(x|z_L) \right] \\
\quad - & \sum_{l=1}^{L} \mathbb{E}_{q_\phi(z_{<l}|x)} \left[ D_{\text{KL}}(q_\phi(z_l|x, z_{<l}) \| p_\theta(z_l|z_{<l})) \right],
\end{split}
\end{equation}
where the first term represents the reconstruction likelihood and the second term contains the KL divergence between the approximate posterior $q_\phi(z_l|x, z_{<l})$ and conditional prior $p_\theta(z_l|z_{<l})$ at each level $l$.

\subsection{SpeakerVAE Adaptation}
To apply NVAE to our 1D speaker embedding data, several adaptations to the original architecture, initially designed for 2D image data, were necessary. Specifically,
we replaced all 2D convolutional operations with 1D equivalents, simplified the model by removing autoregressive normalizing flows while retaining the hierarchical latent structure, implemented robust quantile-based normalization for speaker embeddings, and introduced free-bits regularization \cite{kingma2016improved} with KL coefficient warmup \cite{higgins2017beta} to prevent posterior collapse. We reconfigured the number of hierarchical levels or latent dimensions per level to better suit the dimensionality and inherent complexity of speaker embeddings. The architecture of our SpeakerVAE model is shown in \ref{fig:system}.

During training, we first use the speaker extractor module of the voice conversion models to extract speaker embeddings from training utterances. Then we train the SpeakerVAE model with the ELBO loss. At inference, we 1) sample a novel speaker embedding from SpeakerVAE, with a temperature of 1.0, 2) keep the source utterance's content/prosody representations unchanged, and 3) invoke the voice conversion model to synthesize speech with the source utterance's content/prosody and the generated speaker embedding. Because all modules operate in the same latent domain, no additional alignment or fine-tuning is needed, and Sec.~\ref{sec:res} shows that the resulting conversions retain intelligibility while achieving perceptually distinct, natural-sounding timbres.


\subsection{Voice Conversion Models}
\subsubsection{FACodec}
FACodec \cite{ju2024naturalspeech} is a factorized neural speech codec used as one of our VC models. Unlike traditional residual-VQ codecs, FACodec explicitly decomposes the waveform into four disentangled sub-spaces (content and timbre) and reconstructs speech from these representations with minimal quality loss, making it an ideal architecture for SpeakerVAE. Its strong zero-shot VC capability further allows us to evaluate novel speaker embeddings without any data-specific fine-tuning.

The FACodec system utilizes a 1024-dimensional latent space to represent speaker timbre, learned from scratch on its training data. In the SpeakerVAE training stage, we infer speaker embeddings using the speaker encoder module of a pretrained FAcodec model; these embeddings then serve as the training dataset for SpeakerVAE. We ensured that the corpus from which we inferred these speaker embeddings was part of the pretrained FAcodec model's training dataset to prevent out-of-domain issues that might cause unexpected behavior. During inference, we extract content tokens from source speech and concatenate them with a generated speaker embedding, input to the same pretrained FAcodec model to perform voice conversion. 

We use a widely-accepted unofficial FACodec implementation and checkpoint from \cite{facodecgithub}.


\subsubsection{CosyVoice2}

CosyVoice2 \cite{du2024cosyvoice} is a streaming-capable, zero-shot text-to-speech and voice converion model that factorizes speech generation into three successive modules: a supervised semantic tokenizer, a unified text-speech language model, and a chunk-aware causal flow-matching decoder. Crucially for our work, a speaker embedding is used to provide timbre information during the language model stage. This design allows us to integrate our SpeakerVAE model into the CosyVoice2 system, using its output to replace the original speaker embedding input. Furthermore, CosyVoice2 demonstrates an incredible ability in zero-shot voice conversion, a key feature for this work to synthesize entirely new speakers. The training and inference process for the SpeakerVAE+CosyVoice2 is same as that described for the SpeakerVAE+FAcodec model.

The CosyVoice2 system employs a 192-dimensional CAM++ speaker embedding \cite{wang2023campp}, which is pre-trained on a speaker verification task. For our experiments, we utilize the official implementation and checkpoints provided by the CosyVoice2 authors \cite{cosyvoicegithub} for both the main CosyVoice2 model and its  CAM++ submodule.

\section{Experiment}
\label{sec:exp}
\subsection{Baseline}
We propose a Gaussian Mixture Model (GMM) \cite{reynolds2009gaussian, tacospawn} as a baseline for modeling and generating speaker embeddings. We use the implementation from \cite{sklearn2011}. This model uses $k=12$ components with diagonal covariance matrices. To determine the optimal number of components $k$, we experimented with integer values for $k$ ranging from 3 to 150. For each value, GMM parameters (means, covariance matrices, and mixture weights) were estimated from the training speaker embeddings. To select $k$, we computed the mean squared error (MSE) between the original speaker embeddings and the mean of their most probable GMM component. This MSE metric was chosen over likelihood curves to directly assess how well the GMM components could represent the data points in the embedding space. While the MSE generally decreased with increasing $k$, we selected $k=16$, as this value was identified within an inflection region of the MSE curve, offering a good balance between model complexity and its ability to represent the embeddings. For generation, we then drew 1000 random samples from this fitted GMM. These generated samples were subsequently used as speaker embeddings for the FAcodec and CosyVoice2 vocoder models to synthesize audio waveforms, in the same manner as SpeakerVAE generated embeddings.

\subsection{Dataset}
We experiment on LibriTTS train-clean-100 and LibriTTS train-clean-360 datasets \cite{zen2019libritts}. The split of LibriTTS train-clean-100 offers $\approx$ 460 hours of studio-quality, 24 kHz read speech drawn from public-domain audiobooks. It balances gender with 553 female and 598 male speakers. We extract embeddings for each utterance as one training sample, resulting in 149,715 samples in total.

\subsection{Training Setups}
\subsubsection{Normalization}
To standardize the speaker embeddings while minimizing the influence of outliers, we employ a quantile-based normalization technique \cite{merad2023robust}. This process is applied on a per-feature basis. First, we mitigate outliers by clipping values below the 0.001 quantile and above the 0.999 quantile for each feature. Then each feature is independently scaled and shifted to a target range of [-1, 1] using min-max normalization.

\subsubsection{Model Configurations}
The FACodec system utilizes a 1024-dimensional latent space to represent speaker timbre, learned from scratch using its training data. We use a widely-accepted unofficial FACodec implementation from \cite{facodecgithub}. For the SpeakerVAE model applied to these FACodec embeddings, we configure a hierarchical structure with 2 levels, with each level containing 5 groups, and each group consisting of 20 latent dimensions. Both the encoder and decoder hidden layer sizes are set to 64. 

The CosyVoice2 system employs a 192-dimensional CAM++ speaker embedding \cite{wang2023campp}, which is pre-trained on a speaker verification task. We use the official implementation of CosyVoice2 from \cite{cosyvoicegithub}. When applying SpeakerVAE to these CosyVoice embeddings, we configure it with 2 levels, 3 groups each level, and 8 latent dimensions each group. Similar to the FACodec setup, the encoder and decoder hidden layer sizes are set at 64. 

All SpeakerVAE models are trained on a single Nvidia RTX 4090 GPU for 1000 epochs, using a batch size of 1024.

\subsection{Evaluation Metrics}

\subsubsection{Speaker Generation Quality}
We evaluate speaker generation quality along four key aspects: diversity, coverage, fidelity, and stability. All metrics are computed using speaker embeddings extracted with WavLM-base-plus-sv model \cite{chen2022wavlm}, a state-of-the-art speaker verification system. The cosine similarity between two embeddings is calculated as:
\begin{equation}
\text{cos}(\mathbf{a},\mathbf{b}) = \frac{\mathbf{a} \cdot \mathbf{b}}{\|\mathbf{a}\| \|\mathbf{b}\|}.
\end{equation}
All experiments use $m=1000$ utterances. We construct the following datasets for testing:
\begin{itemize}
\item
\textbf{$GT$ (Ground Truth):} $m$ randomly sampled utterances from the train dataset
\item
\textbf{$GT_{SameSpeaker}$:} $m$ utterances of the training dataset where each one is a randomly selected different utterance with the same speaker for each utterance in GT
\item
\textbf{$S_{Syn}$ (Synthesis with Same Speaker):} $m$ resynthesized utterances, one for each GT utterance by a VC model using the speaker embedding extracted from its corresponding $GT_{SameSpeaker}$ utterance
\item
\textbf{$S_{Recon}$ (Reconstruction of GT):} $m$ reconstructed utterances of the GT utterances
\item
\textbf{$G_{Syn}$ (Synthesis with Generated Speaker):} GT utterances converted to different target pseudo speakers that are generated by our model. 
\end{itemize}

We define the following cosine similarity based metrics for evaluation:

\textbf{Pairwise.} Measures speaker diversity by calculating the average cosine similarity between all possible combinations of distinct utterances from the specified sets, excluding self-comparisons.

\textbf{Corresponding.} Assesses speaker preservation accuracy by calculating the average cosine similarity between matched utterance pairs sharing the same utterance ID.

\textbf{Stability.} Evaluates speaker generation consistency by measuring similarity between speaker embeddings extracted from converted utterances to the same generated speaker but from different source speakers

\textbf{Natural Consistency} Establishes baseline consistency for stability by measuring similarity between different utterances from the same natural speaker, capturing inherent speaker variation.

\subsubsection{Audio Quality}
To quantify intelligibility we report word-error rate (\textbf{WER}) and character-error rate (\textbf{CER}). After synthesizing each utterance, we transcribe it with the open-source Whisper-base ASR model (74 M parameters, multilingual) released by OpenAI \cite{cao2012whisper}. WER is computed as
\begin{equation}
    WER = \frac{S+D+I}{N},
\end{equation}
where $S =$ substitutions, $D =$ deletions, $I =$ insertions, and $N =$ words in the reference transcript. CER applies the same formula at the character level. We implement both metrics with the lightweight JiWER Python toolkit \cite{jiwer2024}, which provides out-of-the-box WER and CER functions as well as a configurable text-normalization pipeline.
For every audio sample we pass Whisper’s hypothesis and the ground-truth transcript to JiWER, average the per-utterance scores over the test sets. 

To rate perceptual quality without human panels, we adopt \textbf{UTMOSv2} \cite{baba2024t05}. UTMOSv2 replaces human listening panels with a feed-forward inference pass: the model takes waveform input, extracts self-supervised features and multi-resolution mel spectrogram and predicts a mean-opinion score through a small fully-connected head. No manual rating loop is involved at evaluation time. 
We report these UTMOSv2-MOS scores alongside intelligibility (WER/CER) to provide an objective proxy for human naturalness judgments in all our experiments.

\section{Results}
\label{sec:res}

\begin{table*}[]
\caption{Speaker generation quality results.}
\label{tab:cosine}
\centering
\begin{tabular}{ccc|cc}
                                                                     & \multicolumn{2}{c|}{FACodec}                & \multicolumn{2}{c}{CosyVoice2}              \\ \cline{2-5} 
                                                                     & \multicolumn{1}{c|}{SpeakerVAE} & GMM       & \multicolumn{1}{c|}{SpeakerVAE} & GMM       \\ \hline
\multicolumn{1}{c|}{$Pairwise(S_{syn}, S_{syn})$, Orginal Diversity}     & \multicolumn{2}{c|}{0.67±0.18}              & \multicolumn{2}{c}{0.65±0.19}               \\
\multicolumn{1}{c|}{Pairwise($G_{syn}$, $G_{syn}$), Generated Diversity} & \multicolumn{1}{c|}{0.74±0.31}  & 0.65±0.19 & \multicolumn{1}{c|}{0.71±0.15}  & 0.73±0.14 \\
\multicolumn{1}{c|}{Pairwise($G_{syn}$, $S_{syn}$), Original Coverage}    & \multicolumn{1}{c|}{0.70±0.15}  & 0.64±0.19 & \multicolumn{1}{c|}{0.67±0.16}  & 0.69±0.16 \\ \hline
\multicolumn{1}{c|}{Pairwise($S_{syn}$, $GT$), Distribution Fidelity}     & \multicolumn{2}{c|}{0.65±0.17}              & \multicolumn{2}{c}{0.67±0.16}               \\
\multicolumn{1}{c|}{Corresponding($S_{syn}$, $GT$), Speaker Fidelity}     & \multicolumn{2}{c|}{0.92±0.04}              & \multicolumn{2}{c}{0.94±0.03}               \\
\multicolumn{1}{c|}{Corresponding($S_{recon}$, $GT$), Speaker Fidelity}   & \multicolumn{2}{c|}{0.93±0.04}              & \multicolumn{2}{c}{0.93±0.04}               \\ \hline
\multicolumn{1}{c|}{Stability}                                       & \multicolumn{1}{c|}{0.85±0.10}  & 0.90±0.06 & \multicolumn{1}{c|}{0.90±0.06}  & 0.86±0.09 \\
\multicolumn{1}{c|}{Natural Consistency}                             & \multicolumn{2}{c|}{0.91±0.05}              & \multicolumn{2}{c}{0.94±0.03}              
\end{tabular}
\end{table*}

\begin{figure}
    \centering
    \includegraphics[width=1.0\linewidth]{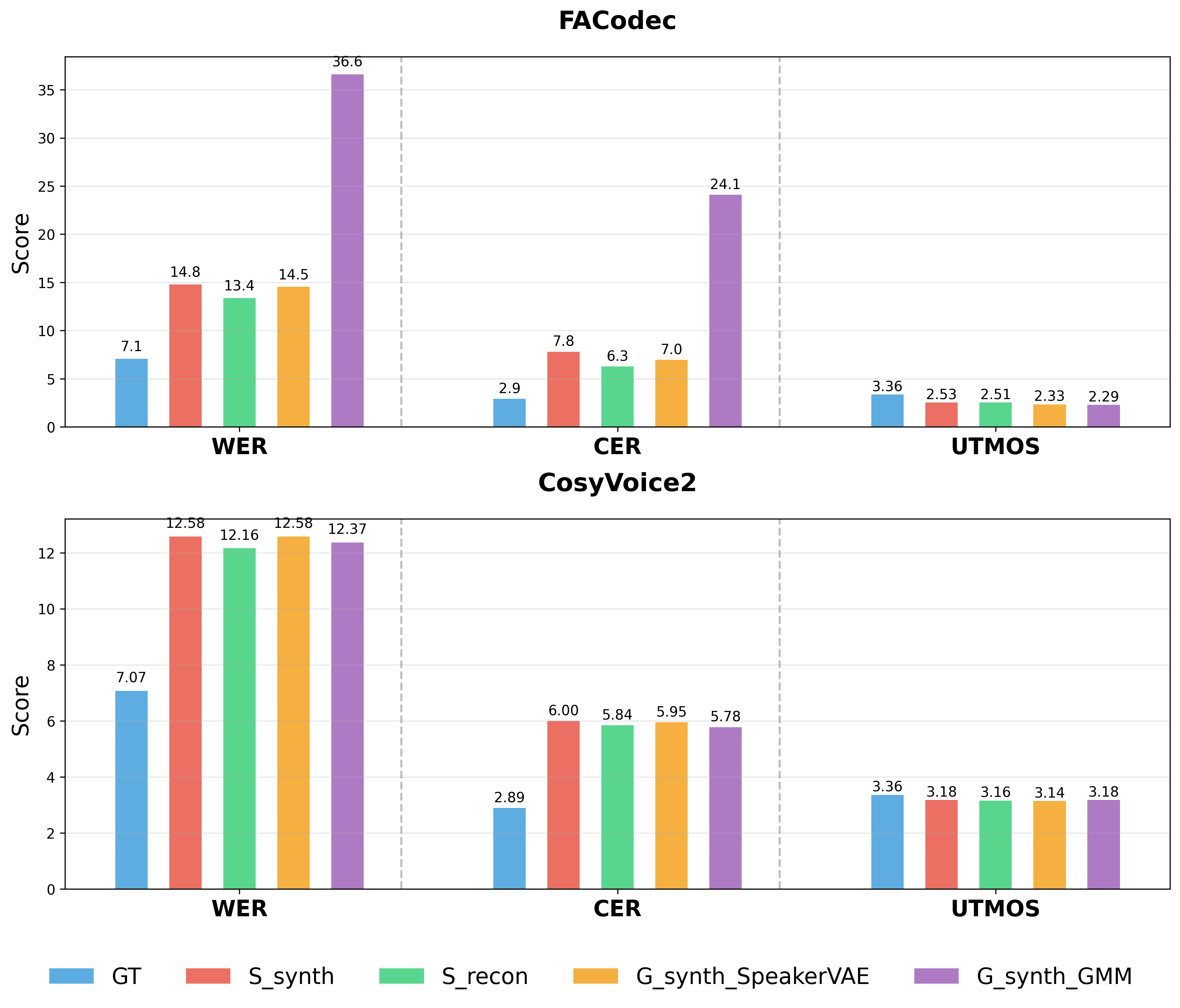}
    \caption{Audio quality metrics.}
    \label{fig:quality}
\end{figure}

\begin{figure}
    \centering
    \includegraphics[width=1.0\linewidth]{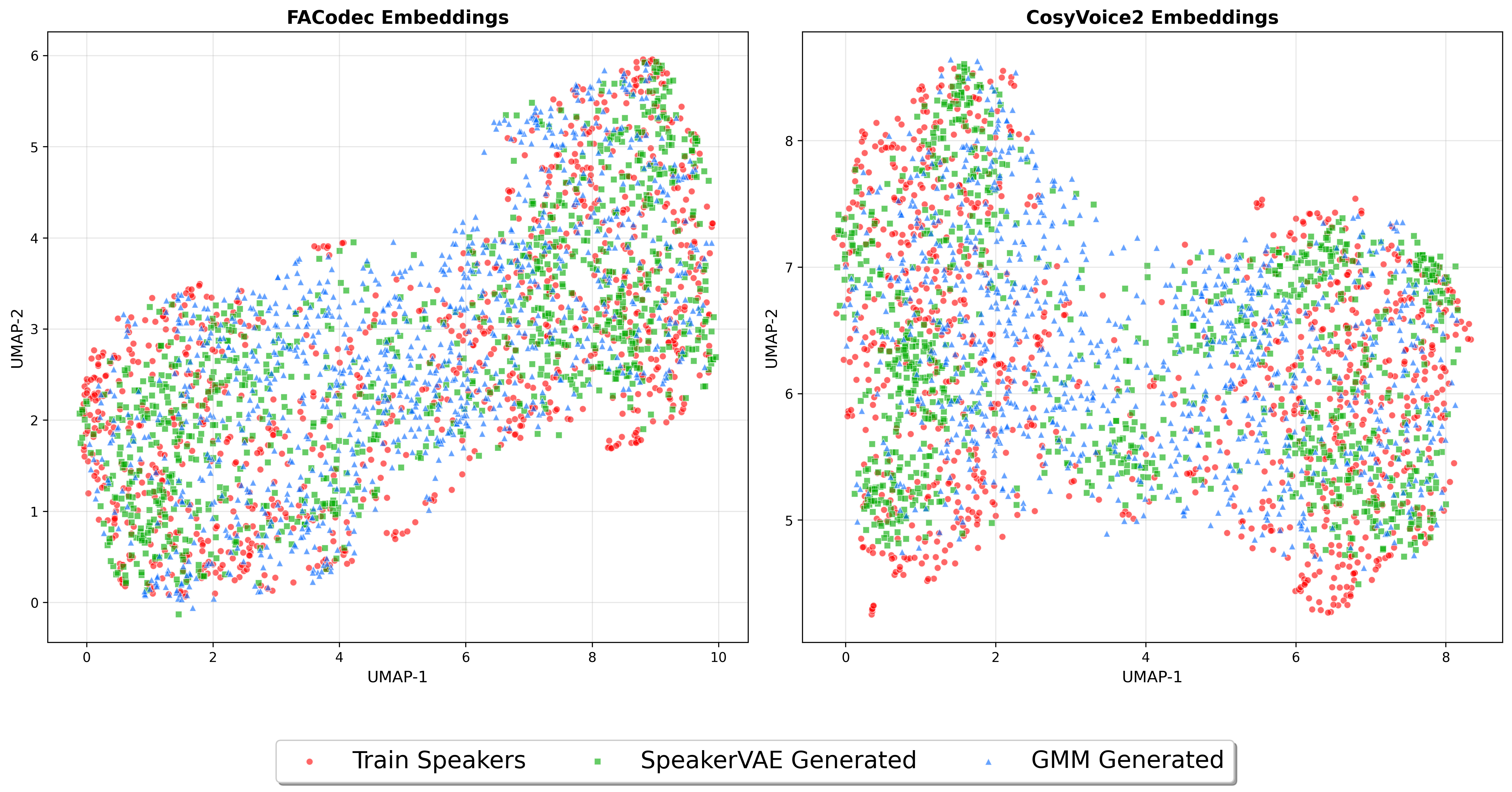}
    \caption{UMAP visualization of the training and generated embeddings.}
    \label{fig:enter-label}
\end{figure}

The audio quality metrics in Fig. \ref{fig:quality} demonstrate that while speech synthesis introduces artifacts that increase WER/CER for both $S_{syn}$ and $S_{recon}$ compared to GT, speech quality remains largely unaffected. For FACodec and CosyVoice2 systems, speaker generation using our method maintains both intelligibility (WER/CER) and perceptual quality (MOS). The exception is the GMM baseline, which degrades FACodec's intelligibility with substantially higher error rates.

The cosine similarity results, detailed in \ref{tab:cosine}, offer a comparative view of speaker generation quality between SpeakerVAE and GMM across the FAcodec and CosyVoice2 systems. 
Regarding diversity and coverage, Pairwise($S_{syn}$, $S_{syn}$) measures internal similarity among resynthesized known speakers (Original Diversity), while Pairwise($G_{syn0}$, $G_{syn0}$) measures model-generated speakers (Generated Diversity). It's important to note that these two diversity metrics are based on pairwise cosine similarities. Therefore, a higher score indicates that the speaker embeddings within that set are more similar to each other, which implies lower diversity.
The $Pairwise(G_{syn0}, S_{syn})$ metric assesses how well generated speakers cover the original range. Ideally, these three metrics would have similar scores, indicating that generated speakers match the original distribution.
For the FACodec system, SpeakerVAE's generated speakers were slightly less diverse  than the resynthesized original set but achieved better coverage, whereas GMM's generated diversity was closer to that of the original set with lower coverage. In the CosyVoice2 setup, both SpeakerVAE and GMM produced speakers that were slightly less diverse than the resynthesized original set. In both VC setups, the three diversity and coverage metrics shows slight variations between the models but indicated reasonable performance. Thus, both SpeakerVAE and GMM demonstrate a good ability to model the original distribution.

The fidelity metrics shows similarity between resynthesized known speakers and their ground truth counterparts. The higher is better. The high scores in the table show both the FACodec and Cosyvoice model can faithfully reconstruct speaker timbre provided by the input speaker embeddings.

Stability measures the consistency of generated speaker identity across different input texts and source speakers. For FACodec, GMM exhibited higher stability than SpeakerVAE , while for CosyVoice2, SpeakerVAE was more stable than GMM. This variation might relate to how effectively each generation method's learned embedding space aligns with the specific characteristics of each VC.  Natural Consistency is use as a reference to the stability measure, since both are cacluted on utterances with same speaker identiy but different texts. The similar number of stability and natural consistency shows the synthesized speech maintains a comparable level of real speech consistency.

\section{Conclusion and Future Work}
\label{sec:conc}

SpeakerVAE enables efficient novel speaker generation for voice conversion by modeling timbre space with a hierarchical VAE. This lightweight, plug-and-play solution requires only VAE training and works across VC systems (FACodec, CosyVoice2). Our evaluation shows it maintains audio quality (WER/CER/UTMOS comparable to original speakers) while generate speakers with original diversity and fidelity. Future work includes developing attribute-controlled (age/gender/accent) speaker generation model using guided latent space sampling.

\label{sec:ack}

\bibliographystyle{IEEEbib}
\bibliography{strings,refs}

\end{document}